\documentclass[superscriptaddress]{revtex4-2}
\usepackage{graphicx}
\usepackage{subfig}
\usepackage{color}
\usepackage{amsmath}
\usepackage{amssymb}
\usepackage{bm}
\usepackage{slashed}
\usepackage{epsfig}
\usepackage{amsfonts}
\usepackage{epstopdf}
\usepackage{textcomp}
\usepackage{enumitem}
\usepackage{subfig}
\usepackage{hyperref}
\usepackage{cancel}
\usepackage{booktabs}
\usepackage{appendix}
\usepackage{amsthm}
\usepackage{mathrsfs}

\begin{document}

\title{Study of the rare decay $J/\psi \to 2\gamma+hadrons$ at the BESIII}
\author{Feng Zhang}
\email{zhangfeng@ihep.ac.cn}
\author{Jian-Xiong Wang}
\email{jxwang@ihep.ac.cn}
\affiliation{
	Institute of High Energy Physics (IHEP), Chinese Academy of Sciences (CAS),
	19B Yuquan Road, Shijingshan District, Beijing, 100049, China}
\affiliation{University of Chinese Academy of Sciences (UCAS), Chinese Academy of Sciences (CAS),
	19A Yuquan Road, Shijingshan District, Beijing, 100049, China}

\date{\today}

\begin{abstract}
Two-photon radiative decay process $J/\psi \to 2\gamma+hadrons$ 
is studied, and the main contribution processes $J/\psi \to 2\gamma + g g g$ and $J/\psi \to 2\gamma + q \bar{q}$ are calculated.
With the specific condition at the BESIII, this rare decay process and the main background process $e^{+} e^{-} \to \gamma \gamma + hadrons (q \bar{q})$ are investigated.  
The results show that the ratio of signal to background can reach 1.24 with the optimized selection criteria at the BESIII. 
In addition, a few distributions
of the signal and background are presented. All the results show that the signal is large enough to be measured in the experiment.

\noindent\\
{\bf Keywords:}  rare decay of $J/\psi$, branch ratio,  BESIII
\end{abstract}

\maketitle
\allowdisplaybreaks

\section{Introduction}
As the number of $J/\psi$ events at the BESIII reaches 10 billion~\cite{BESIII:2012pbg,BESIII:2016kpv,BESIII:2021cxx}, a few rare decays of $J/\psi$ may be 
measured at the BESIII experiment, such as the prediction and observation of the four-lepton decay process~\cite{Chen:2020bju,BESIII:2021ocn}. 
In comparison with the clean property of lepton decay process, the rare hadron decay process is harder to find in experiments 
because of final-state quarks and gluon hadronization, small signal and complicated background.  The one-photon radiative decay channels $J/\psi \to \gamma+hadrons$ have been 
well studied~\cite{ParticleDataGroup:2020ssz},  while two-photon channels $J/\psi \to 2\gamma+hadrons$ are much more difficult
to find.  A few cases were measured, in which one of the photons is from internal hadronic resonances, such as $\eta(1405/1475)$~\cite{MARK-III:1989jot,
BES:2004pec,BESIII:2018dim}. And the case with two photons directly radiated from $J/\psi$ decay before final-state quarks and gluon hadronization has not been found. 

It is easy to see that the channels of two photons radiated directly from $J/\psi$ decay are existed but small. With the big number of $J/\psi$ 
events at the BESIII, there is a chance to measure the rare hadron decay process $J/\psi \to  2\gamma + hadrons$. 
But the detailed information of the signal and background are needed to find out optimization selection criteria in 
the measurement.  

It is well known that hadron can not be treated directly in the perturbative calculation due to 
the non-perturbative property of Quantum Chromodynamics  (QCD), and a factorization scheme has to be introduced 
to factorize the perturbative part and non-perturbative part, where non-perturbative parameters could be fixed 
from experimental measurements and the 
perturbative part can be calculated in perturbative expansion of QCD and other small parameter.   
For the signal of $J/\psi$ rare decay, a factorization scheme is applied in our calculation. 
The scheme is called non-relativistic Quantum Chromodynamics (NRQCD)~\cite{Bodwin:1994jh} which were successfully applied 
in many cases.  

In this work, we calculate the signal and background in detail, optimize the selection criteria to suppress the background and present 
numerical estimate results, which are useful for future experimental measurement. 

\section{The signal}
\begin{figure}[ht]
	\centering
	\includegraphics[width=15cm]{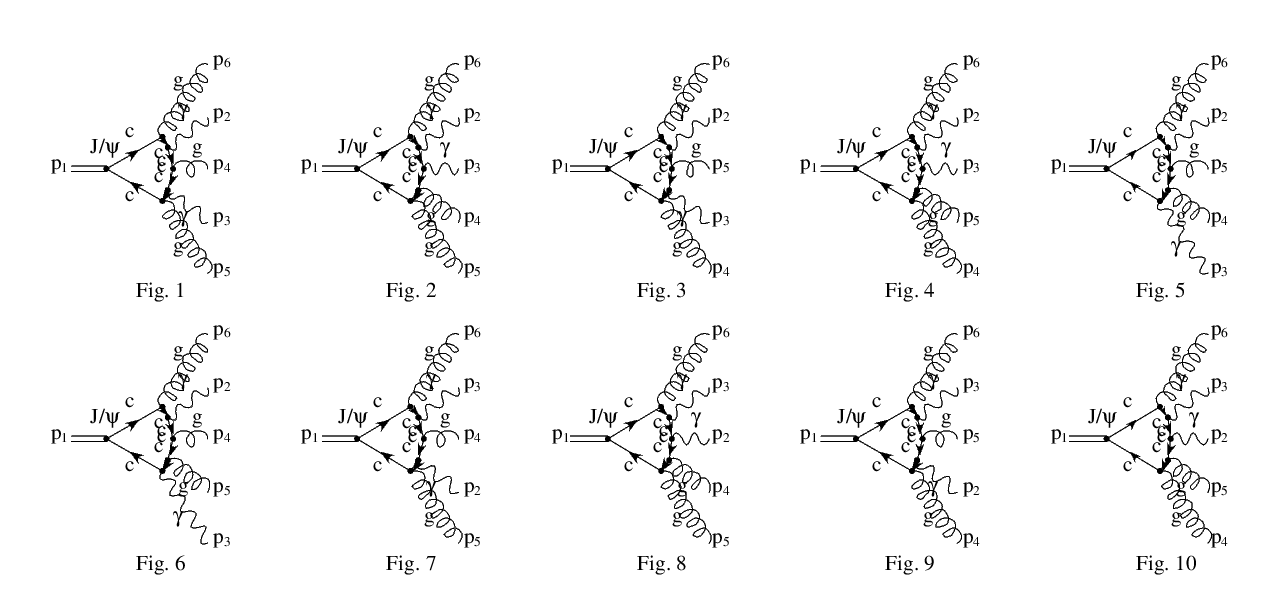}
	\caption{Part of the Feynman diagrams for the process $J/\psi \to \gamma \gamma + g g g$}
	\label{ppggg}
\end{figure}

\begin{figure}[ht]
	\centering
	\includegraphics[width=16cm]{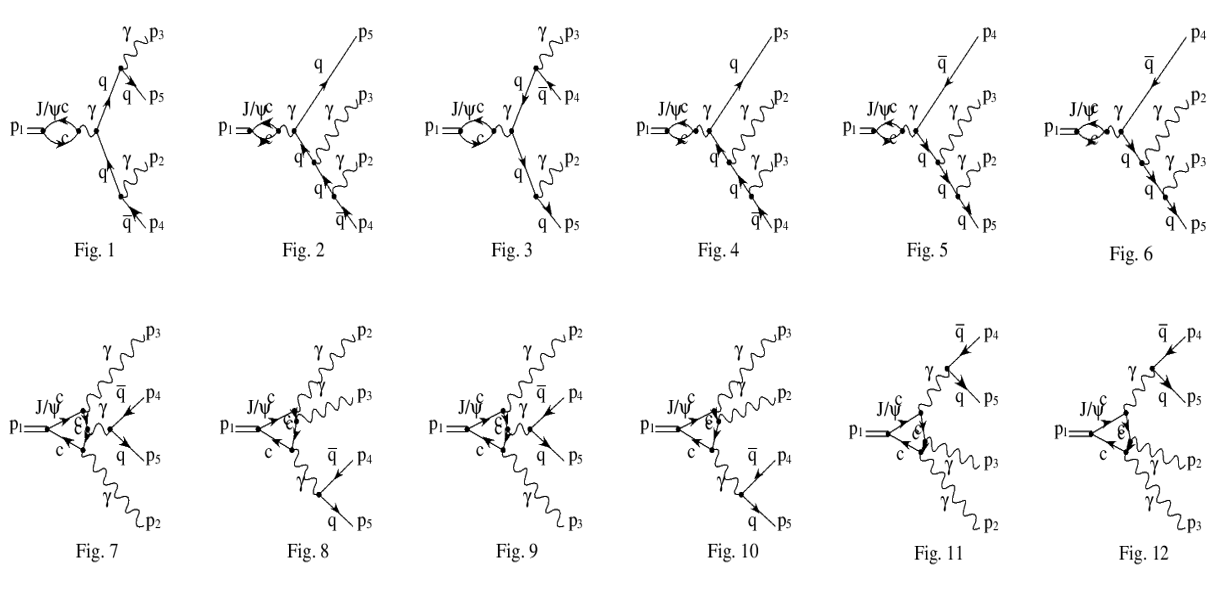}
	\caption{The Feynman diagrams for the process $J/\psi \to \gamma \gamma + q \bar{q}$ }
	\label{jppqq}
\end{figure}

The leading contribution to the two-photon radiative decay  $J/\psi \to  2\gamma + hadrons$ is from the processes $J/\psi \to 2 \gamma + ggg$ and $J/\psi \to 2 \gamma + q \bar{q}$, where q is light quarks (u, d and s flavors). 
The process $J/\psi \to 2 \gamma + ggg$ contains 120 Feynman diagrams, and part of them is shown in FIG.~\ref{ppggg}.
The decay width equals
\begin{equation}
	\Gamma(J/\psi \to \gamma \gamma + g g g)=\frac{1}{2M_{J/\psi}}\int d\Pi_5\frac{1}{3}\sum_{\text{polarization}}|\mathcal{M}(
	J/\psi \to \gamma \gamma + g g g)|^2,
	\label{yyggge}
\end{equation}
where $d\Pi_5$ is the five-body phase space with possible symmetry factors due to identical final-state particles, and the polarization summation 
contains two transverse and one longitudinal states of $J/\psi$ and polarization states of all final-state particles. The process $J/\psi \to 2 \gamma + q \bar{q}$ contains 12 Feynman diagrams, which are shown in FIG.~\ref{jppqq}.
The decay width equals
\begin{equation}
\Gamma(J/\psi \to \gamma \gamma + q \bar{q})_{fixed}=\frac{1}{2M_{J/\psi}}\int d\Pi_{4,fixed}\frac{1}{3}\sum_{\text{polarization}}|\mathcal{M}(
J/\psi \to \gamma \gamma + q \bar{q})|^2,
\label{zyyggge}
\end{equation}
where $d\Pi_4$ is the four-body phase space with possible symmetry factors due to identical final-state particles, and  
there is obvious infrared and collinear divergence problem to define two photon associated with hadron production both in 
theoretical calculation and experimental measurement. Therefore, thereafter a subscript "$_{fixed}$" will put on 
all the decay width which is needed to declare a infrared safety cut condition in theoretical calculation 
and experimental measurement.  

The branching ratio is defined as:
\begin{equation}
\mathcal{B}(J/\psi \to \gamma \gamma + ggg(q\bar{q}))_{fixed}=\frac{\Gamma(J/\psi \to \gamma \gamma + ggg(q\bar{q}))_{fixed}}{\Gamma_{\text{total}}}
\label{zeq1},
\end{equation}
to estimate the $\mathcal{B}(J/\psi \to \gamma \gamma + hadrons)_{fixed}$ in our calculation. However,
the absolute width is of large uncertainties in the calculation.
To minimize the uncertainties from the running strong interaction coupling constant, wave function of $J/\psi$ at origin and higher order QCD correction, the branching ratio for these two contribution processes shall be obtained by different bridge processes.

\begin{figure}[ht]
	\centering
	\includegraphics[width=17cm]{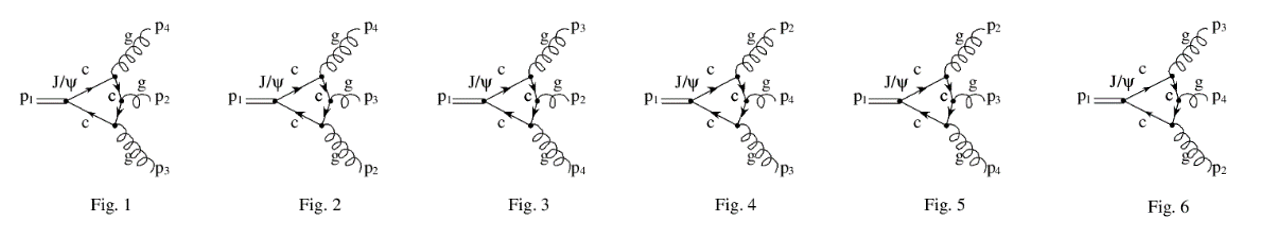}
	\caption{The Feynman diagrams for the process $J/\psi \to g g g$ }
	\label{ggg}
\end{figure}

For the process $J/\psi \to 2 \gamma + ggg$, the branching ratio equals
\begin{equation}
\mathcal{B}(J/\psi \to \gamma \gamma + ggg)_{fixed}=\frac{\Gamma(J/\psi \to \gamma \gamma +ggg)_{fixed}}
	{\Gamma(J/\psi \to g g g)_{fixed}} \mathcal{B}(J/\psi \to hadrons(ggg)).
\label{eq1}
\end{equation}
It is just proportional to the electromagnetic fine structure constant $\alpha^2$. In the numerical calculation $\alpha=1/128$ is used, and the bridge branching ratio is~\cite{ParticleDataGroup:2020ssz,CLEO:2008gct}
\begin{equation}
\mathcal{B}(J/\psi \to hadron(ggg))=(64.1 \pm 1.0)\%.
\label{ratio}
\end{equation}
The leading order of the bridge process $J/\psi \to g g g$ contains 6 Feynman diagrams shown in FIG.~\ref{ggg}
and the decay width equals
\begin{equation}
\Gamma(J/\psi \to g g g) 
=\frac{1}{2M_{J/\psi}}\int d\Pi_3\frac{1}{3}\sum_{\text{polarization}}|\mathcal{M}(J/\psi \to  g g g)|^2,
\label{ggge}
\end{equation}
where $d\Pi_3$ is the three-body phase space with possible symmetry factors due to identical final-state particles. 

\begin{figure}[ht]
	\centering
	\includegraphics[width=4cm]{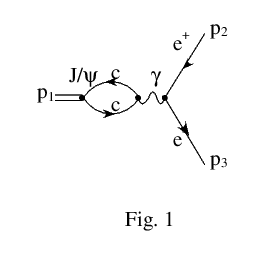}
	\caption{The Feynman diagram for the process $J/\psi \to e^+e^-$ }
	\label{ee}
\end{figure}

For the process $J/\psi \to 2 \gamma + q \bar{q}$, the branching ratio equals
\begin{equation}
\mathcal{B}(J/\psi \to \gamma \gamma + q \bar{q})_{fixed}=\frac{\Gamma(J/\psi \to \gamma \gamma +q \bar{q})_{fixed}}
	{\Gamma(J/\psi \to e^+e^-)_{fixed}} \mathcal{B}(J/\psi \to e^+e^-).
\label{eq2}
\end{equation}
It is also proportional to the electromagnetic fine structure constant $\alpha^2$, and the bridge branching ratio is~\cite{ParticleDataGroup:2020ssz}
\begin{equation}
\mathcal{B}(J/\psi \to e^+ e^-)=(5.971 \pm 0.032)\%.
\label{ratio2}
\end{equation}
The leading order of the bridge process $J/\psi \to e^+e^-$ contains 1 Feynman diagram shown in FIG.~\ref{ee}
and the decay width equals
\begin{equation}
\Gamma(J/\psi \to e^+e^-) 
=\frac{1}{2M_{J/\psi}}\int d\Pi_2\frac{1}{3}\sum_{\text{polarization}}|\mathcal{M}(J/\psi \to  e^+e^-)|^2,
\label{eee}
\end{equation}
where $d\Pi_2$ is the two-body phase space. 
It should be mentioned that the dominate contribution of $J/\psi \to \gamma \gamma + q \bar{q}$ is from Fig.1-6 in 
FIG.~\ref{jppqq}.  Hence the higher order QCD correction effects could be minimized in Eq.(\ref{eq2}).

With Eq.(\ref{eq1})(\ref{eq2}), the branching ratio for the rare decay of $J/\psi$ are obtained as 
\begin{equation}
\begin{split}
	\mathcal{B}(J/\psi \to \gamma \gamma +g g g)|_{fixed}=(1.20 \pm 0.02) \times 10^{-6},	 \\
	\mathcal{B}(J/\psi \to \gamma \gamma + q \bar{q})|_{fixed}=(2.08 \pm 0.01) \times 10^{-5}, \\ 	
	\mathcal{B}(J/\psi \to \gamma \gamma + hadrons)|_{fixed}=(2.20 \pm 0.01) \times 10^{-5},	
\end{split}
\end{equation}
where the infrared safety cut condition is fixed as: the energy for each photon is $E_\gamma>0.05$ GeV 
and the angle between photons and gluons (quarks) is $cos\theta_{\gamma g, \gamma q, \gamma \bar{q}}<0.95$, and the hadronization requirements:
\begin{equation}
\begin{split}
m_{ggg}^2 &\equiv (\sum_{\text{all gluon}}p)^2 > (2 \times 0.13 \text{ GeV})^2, \\
m_{q\bar{q}}^2 &\equiv (p_q+p_{\bar{q}})^2 > 
\begin{cases}
(2 \times 0.13 \text{ GeV})^2, & \text{for $u$ or $d$ quarks} \\ 
(2 \times 0.49 \text{ GeV})^2, & \text{for $s$ quark} 
\end{cases}
\end{split}
\label{ee1}
\end{equation}
where $p$ is the 4-momentum of gluon, 0.13 GeV is the mass of pion, and 0.49 GeV is the mass of kaon. The calculation of all decay widths for $\Gamma(J/\psi \to g g g)$, $\Gamma(J/\psi \to \gamma \gamma + g g g)$, 
$\Gamma(J/\psi \to \gamma \gamma + q \bar{q})$ and $\Gamma(J/\psi \to e^+e^-)$ are carried out based on 
the leading-order NRQCD contribution by using FDC package~\cite{Wang:2004du}. 

There are two ways to measure the two-photon radiative decay channel of $J/\psi$. One is to utilize 
$J/\psi$ from $\psi(3686)\rightarrow J/\psi +\pi^+ + \pi^-$, 
for which there are $N_{\psi(3686)}=(4.481\pm0.029)\times 10^8$ $\psi(3686)$ events, just as the recent  
observation of $J/\psi$ decay to $e^+e^-e^+ e^-$ and $e^+ e^- \mu^+ \mu^-$ in the BESIII~\cite{BESIII:2021ocn}. 
And an estimation of the two-photon radiative decay event number is 
\begin{eqnarray}
	&N(J/\psi \to \gamma \gamma + hadrons)_{fixed}=N_{\psi(3686)} \times \mathcal{B}(\psi(3686) \to J/\psi \pi^+ \pi^-) 
	\times \mathcal{B}(J/\psi \to \gamma \gamma +hadrons)_{fixed}=3418, 
\end{eqnarray}
where $\mathcal{B}(\psi(3686) \to J/\psi \pi^+ \pi^-)=(34.68\pm 0.30)\%$ is used~\cite{ParticleDataGroup:2020ssz}.
There is no large background from theoretical consideration, but there are a few background in the experimental 
measurement. 

Another is to search for the two-photon radiative decay channel of $J/\psi$ from the data sample of $J/\psi$ at the BESIII.
In this case, the results of the branching ratio can not be applied directly since 
there is some special difference to be considered. 
For the decay width Eq.(\ref{yyggge}),   one needs to notice that the polarization summation for initial-state $J/\psi$ 
should only contain the transverse polarization together with detector coverage limitation
because almost only transverse polarization $J/\psi$ can be produced in the electron-positron collider, 
while the longitudinal polarization $J/\psi$ can be ignored. In this condition, for example, the decay width for the process $J/\psi \to \gamma \gamma + g g g$ equals
\begin{equation}
\Gamma(J/\psi \to \gamma \gamma + g g g)=\frac{1}{2M_{J/\psi}}\int d\Pi_5\frac{1}{2}\sum_{\text{polarization}}|\mathcal{M}(J/\psi \to \gamma \gamma + g g g)|^2.
\end{equation}
In addition, according to the $J/\psi$ detector condition at the BESIII~\cite{BESIII:2021cxx}, the energy cut for each photon is $0.05$ GeV,
and the polar angle cut for each final-state particle is $|cos\theta|<0.93$, where the polar angle means the angle between the particle and the beam direction. 
In experiment, gluon can not be detected directly, and the detectable particles are hadrons, which requires $m_{ggg}^2> (2 \times 0.13 \text{ GeV})^2$. Although Eq.(\ref{ee1}) is trivial for the process $J/\psi \to  g g g$, 
it really cut the phase space for $J/\psi \to \gamma \gamma + g g g$. 
All the above considerations are also taken into consideration for  $J/\psi \to \gamma \gamma + q \bar{q}$. 
In the following calculation,
$N_{\text{total}}=(1.0087\pm0.0044) \times 10^{10}$ is the event number of $J/\psi$ at the BESIII~\cite{BESIII:2021cxx}  
with the corresponding integrated luminosity $\mathcal{L}=3083 pb^{-1}$ and the center of mass energy $\sqrt{s}=M_{J/\psi}=3.097 \text{GeV}$.

\section{The main background}
\begin{figure}[ht]
	\centering
	\includegraphics[width=15cm]{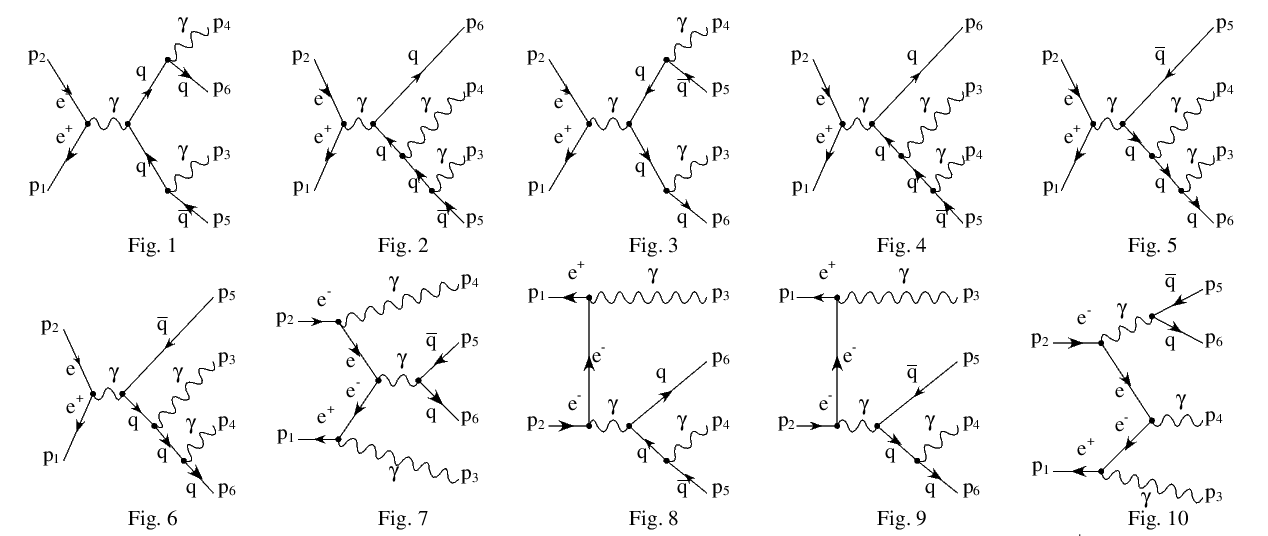}
	\caption{Part of the Feynman diagrams for the process $e^{+} e^{-} \to \gamma \gamma + q \bar{q}$}.
	\label{ppqq}
\end{figure}

To research on rare process, a detailed background study is very important. For the signal of the process $J/\psi \to \gamma \gamma + hadrons$, there could 
be a few source for the background, such as final state radiation, resonance radiative decay, etc.  The main background is the process
$e^{+} e^{-} \to \gamma \gamma + q \bar{q}$, 
where q is light quarks (u, d and s flavors), with 20 Feynman diagrams, and part of them shown in FIG.~\ref{ppqq}. 
And the cross section is 
\begin{equation}
\sigma(e^{+} e^{-} \to \gamma \gamma + q \bar{q})=\frac{1}{2s}\int d\Pi_4\frac{1}{4}\sum_{\text{polarization}}|\mathcal{M}(e^{+} e^{-} \to \gamma \gamma + q \bar{q})|^2,
\end{equation}
where $d\Pi_4$ is the four-body phase space with possible symmetry factors due to identical final-state particles. 
The calculation of the cross section is carried on by using FDC package~\cite{Wang:2004du}. 
Then the event number equals
\begin{equation}
N(e^{+} e^{-} \to \gamma \gamma + hadrons)=\sigma(e^{+} e^{-} \to \gamma \gamma + q \bar{q})*\mathcal{L}.
\end{equation}

\section{The optimization of selection criteria}  
All the following optimizations are based on the event sample data generated from the signal and background calculation with the corresponding event number. 
There are a few selection criteria for the signal $J/\psi \to \gamma \gamma +hadrons$ and background $e^{+} e^{-} \to \gamma \gamma + hadrons (q \bar{q})$ that need to consider, which are shown in TABLE~\ref{tab:one}.
\begin{table}[!htb]
	\begin{tabular}{ccc} 
		\hline
		\hline
		Process & $J/\psi \to \gamma \gamma + hadrons$ & $e^{+} e^{-} \to \gamma \gamma + hadrons$ \\
		\hline	
	    1. The cut of energy for each photon &    $E_\gamma>E_{\gamma, cut}$  & \quad  $E_\gamma>E_{\gamma, cut}$   \\
        2. The cut of polar angle for each final-state particle &$|cos\theta|<cos\theta_{cut}$ &\quad $|cos\theta|<cos\theta_{cut}$\\
        3. The cut of invariant mass for hadrons production & $m_{ggg, q\bar{q}}^2 > s_{cut}$& \quad$m_{q\bar{q}}^2 > s_{cut}$\\	
        4. The cut of angle between photons and gluons (quarks) &  $cos\theta_{\gamma g, \gamma q, \gamma \bar{q}}<cos\theta_{\gamma g, \gamma q, cut}$ &  
        \quad $cos\theta_{\gamma q, \gamma \bar{q}}<cos\theta_{\gamma g, \gamma q, cut}$\\
		\hline
		\hline
	\end{tabular}
    \caption[]{The selection criteria for the processes $J/\psi \to \gamma \gamma + hadrons$ and $e^{+} e^{-} \to \gamma \gamma + hadrons$}
    \label{tab:one}	
\end{table}

The basic cut values in TABLE~\ref{tab:one} equal:
\begin{equation}
E_{\gamma, cut}=0.05 \text{ GeV},\quad
cos\theta_{cut}=0.93,\quad
s_{cut}=
	\begin{cases}
	(2 \times 0.13 \text{ GeV})^2, & \text{for $u$ or $d$ quarks} \\ 
	(2 \times 0.49 \text{ GeV})^2, & \text{for $s$ quark} 
	\end{cases},\quad
cos\theta_{\gamma g, \gamma q, cut}=0.95,
\label{eq13}
\end{equation}
where the word 'basic' means the range of these values refers to the maximum range of event selection.
The basic criterion 1 and 2 are $J/\psi$ selection criteria at the BESIII detector~\cite{BESIII:2021cxx}; 
the basic criterion 3 is for hadron production; 
the basic criterion 4 is for avoiding the problem of collinear divergence in the processes $e^{+} e^{-} \to \gamma \gamma + q \bar{q}$ and $J/\psi \to \gamma \gamma +q \bar{q}$.  

With these basic selection criteria, the branching ratio for the rare decay $J/\psi \to \gamma \gamma + g g g$ is obtained as 
\begin{equation}
\begin{split}
\mathcal{B}(J/\psi \to \gamma \gamma +g g g)=(1.01 \pm 0.02) \times 10^{-6},	\\
\mathcal{B}(J/\psi \to \gamma \gamma + q \bar{q})=(1.74 \pm 0.01) \times 10^{-5}, \\
\mathcal{B}(J/\psi \to \gamma \gamma + hadrons)=(1.84 \pm 0.01) \times 10^{-5},	
\end{split} 	
\end{equation}
and the numbers of events corresponding to $N_{\text{total}}=(1.0087\pm0.0044) \times 10^{10}$ (the event number 
of $J/\psi$ at the BESIII) are shown as
\begin{equation}
	\begin{split}	
		N(J/\psi \to \gamma \gamma u \bar{u})=170279,&
		\quad
		N(J/\psi \to \gamma \gamma d \bar{d})=2695,
		\quad 
		N(J/\psi \to \gamma \gamma s \bar{s})=2465, \\
		N(J/\psi \to \gamma \gamma g g g)=10180, &
		\quad 
		N(J/\psi \to \gamma \gamma q \bar{q})=175439, \\
		N(e^{+} e^{-} \to \gamma \gamma u \bar{u})=157382,&
		\quad
		N(e^{+} e^{-} \to \gamma \gamma d \bar{d})=25917,
		\quad 
		N(e^{+} e^{-} \to \gamma \gamma s \bar{s})=17824, \\
		N(e^{+} e^{-} \to \gamma \gamma + hadrons)=201123,& 
		\quad 		
		N(J/\psi \to \gamma \gamma + hadrons )=185619,
		\quad
		R=92.3\%,
		\label{y0}
	\end{split}
\end{equation} 
where 
\begin{equation}
\begin{split}
N(e^{+} e^{-} \to \gamma \gamma + hadrons)=N(e^{+} e^{-} \to \gamma \gamma u \bar{u})+N(e^{+} e^{-} \to \gamma \gamma d \bar{d})+N(e^{+} e^{-} \to \gamma \gamma s \bar{s}), \\
N(J/\psi \to \gamma \gamma + q \bar{q})=N(J/\psi \to \gamma \gamma u \bar{u})+N(J/\psi \to \gamma \gamma d \bar{d})+N(J/\psi \to \gamma \gamma s \bar{s}), \\
N(J/\psi \to \gamma \gamma + hadrons)=N(J/\psi \to \gamma \gamma ggg)+N(J/\psi \to \gamma \gamma q \bar{q}), 
\end{split}
\end{equation}
and R is a ratio: 
\begin{equation}
	R \equiv \frac{N(J/\psi \to \gamma \gamma + hadrons)}{N(e^{+} e^{-} \to \gamma \gamma + hadrons)}.
	\label{R}
\end{equation}

\begin{figure}[ht]
        \centering
        \subfloat{
                \includegraphics[width=8cm]{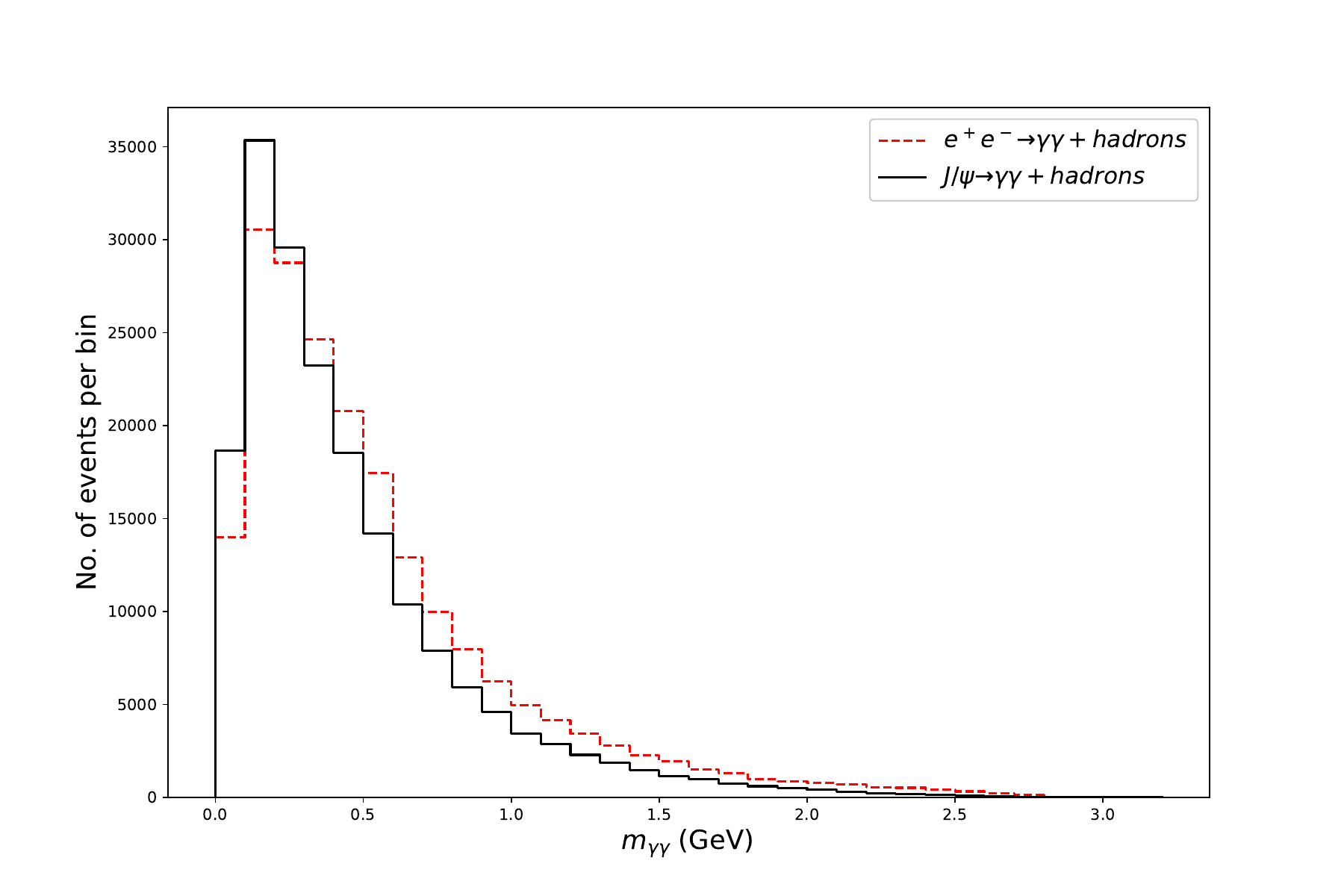}
        }
        \subfloat{
                \includegraphics[width=8cm]{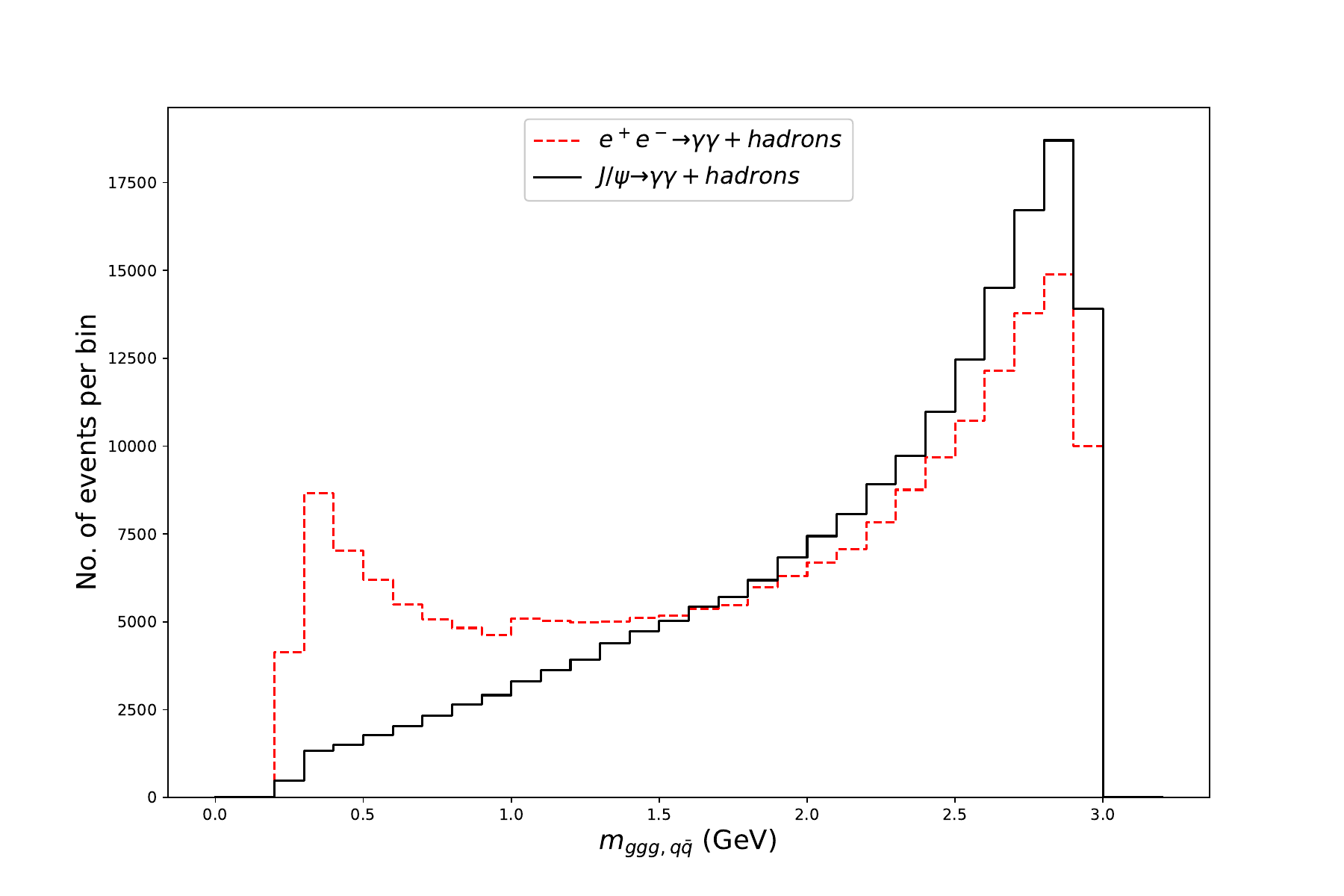}
        }\\
        \subfloat{
                \includegraphics[width=8cm]{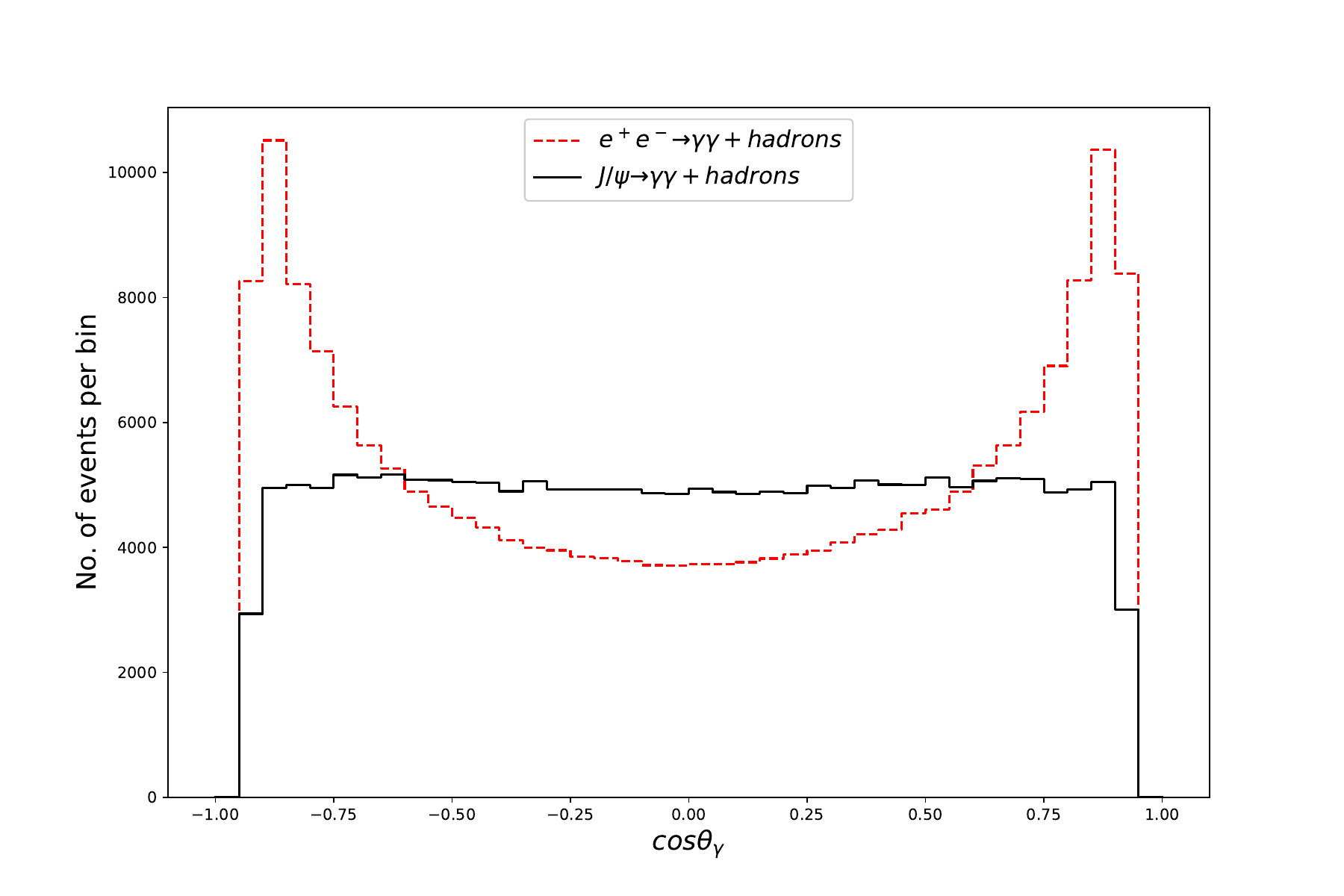}
        }
        \subfloat{
        	\includegraphics[width=8cm]{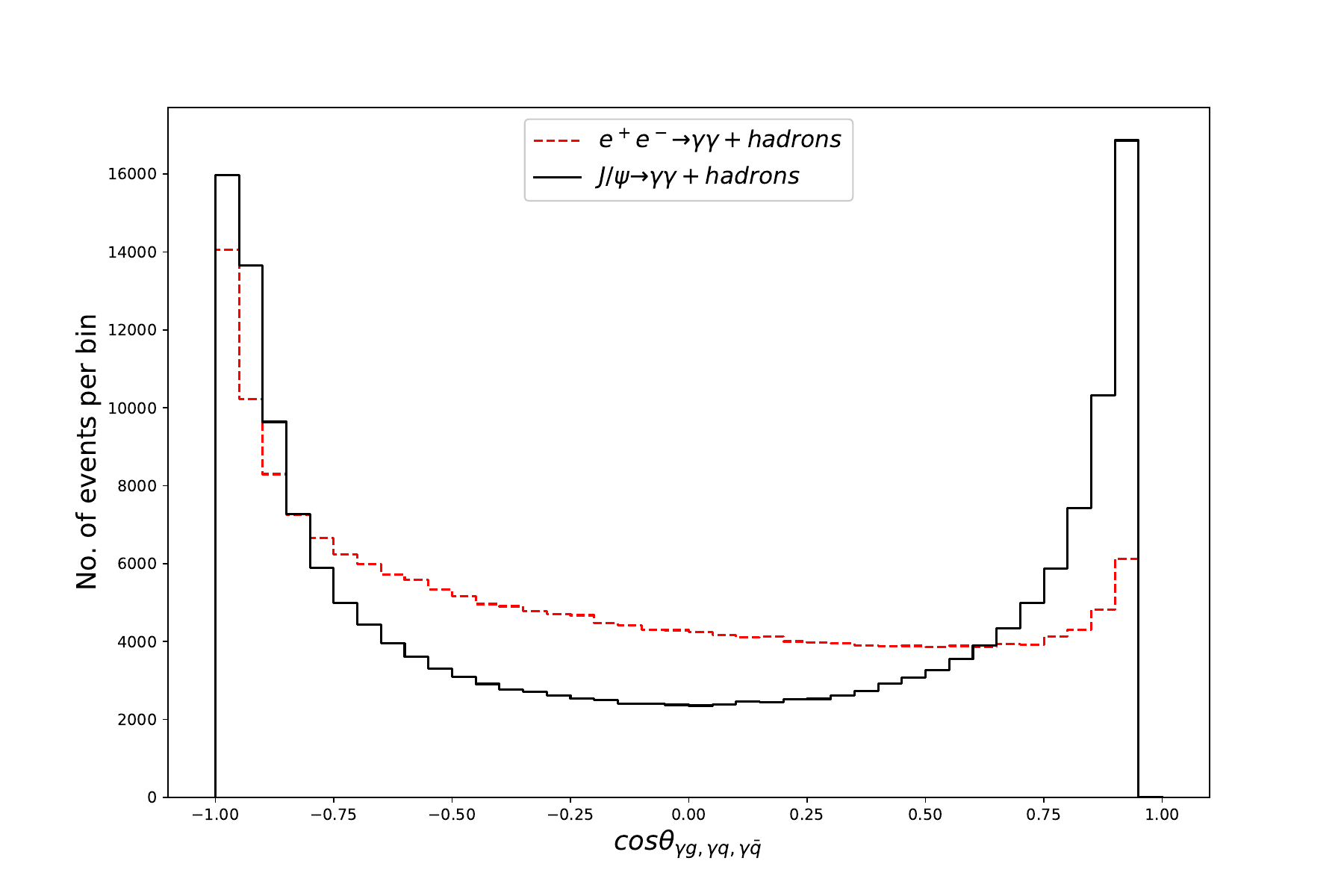}
        }\\
        \subfloat{
        	\includegraphics[width=8cm]{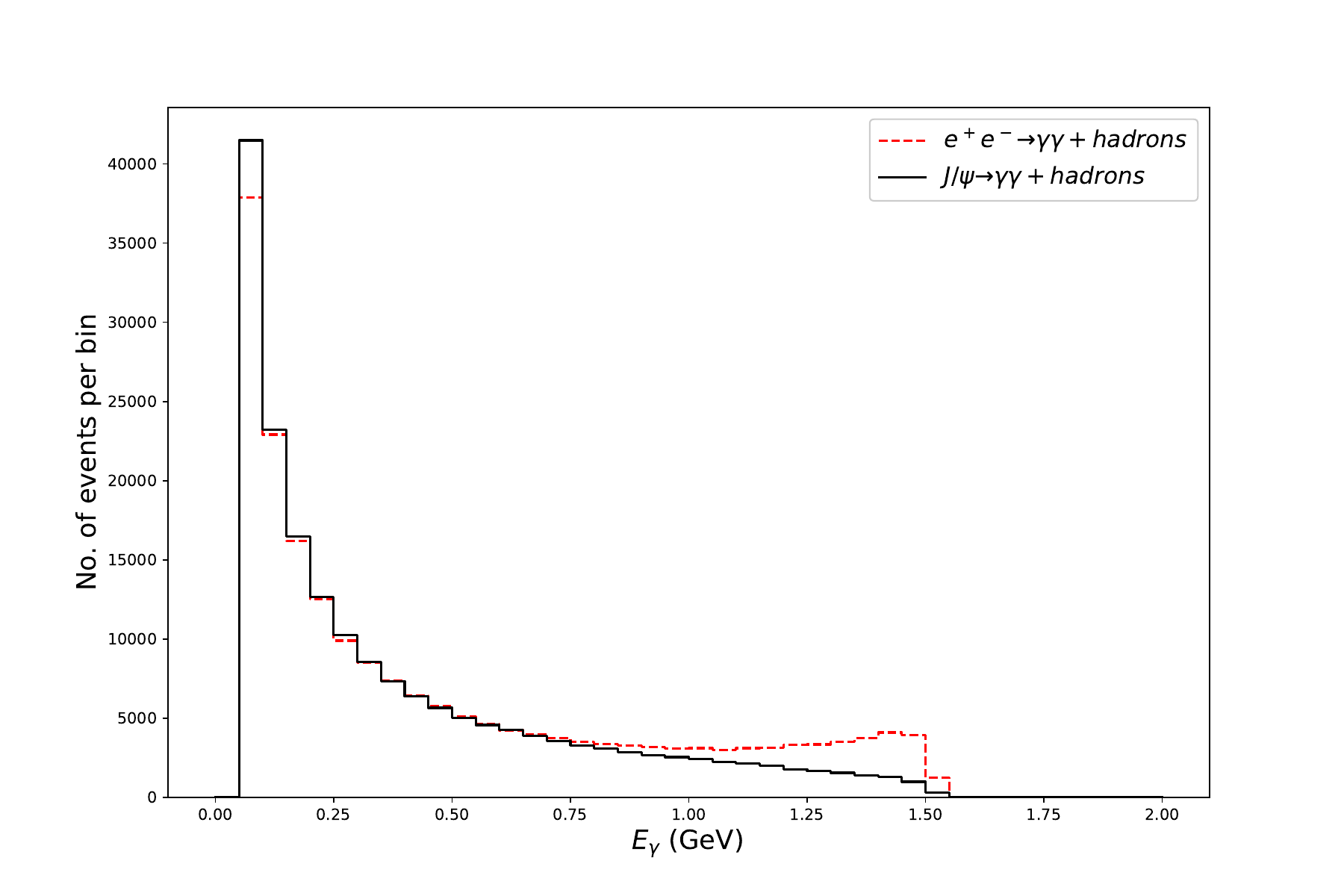}
        }
        \caption{The distributions of $J/\psi \to \gamma \gamma + hadrons$ and $e^{+} e^{-} \to \gamma \gamma + hadrons$ with the basic selection criteria.
        The bin widths are 0.1 GeV for $m_{\gamma\gamma}$ and $m_{ggg, q \bar{q}}$, 0.05 GeV for $E_\gamma$, and 0.05 for $cos\theta_\gamma$ and $cos\theta_{\gamma g, \gamma q, \gamma \bar{q}}$.  }
        \label{disy1}
\end{figure}

The distributions of two-photon invariant mass $m_{\gamma\gamma}$ and the hadron invariant mass $m_{ggg, q \bar{q}}$ 
for the signal $J/\psi \to \gamma \gamma + hadrons$ and the background process $e^{+} e^{-} \to \gamma \gamma + hadrons$ are shown in FIG.~\ref{disy1}.
The distributions of $cos\theta_\gamma$, $E_\gamma$ and $cos\theta_{\gamma g, \gamma q, \gamma \bar{q}}$ for signal and background are also presented in FIG.~\ref{disy1}, 
where $cos\theta_\gamma$ and $E_\gamma$ are the polar angle and energy of photons and each photon is counted as 0.5 because of two indistinguishable photons in each event, while $cos\theta_{\gamma g, \gamma q, \gamma \bar{q}}$ is the angle between photons and gluons (or quarks) and each pair is counted as 1/6 (or 1/4) for the same reason. 
In the distribution plots,  the bin widths are 0.1 GeV for $m_{\gamma\gamma}$ and $m_{ggg, q \bar{q}}$, 0.05 GeV for $E_\gamma$, and 0.05 for $cos\theta_\gamma$ and $cos\theta_{\gamma g, \gamma q, \gamma \bar{q}}$.
Through the ratio R in results~Eq.(\ref{y0}) and the distributions shown in FIG.~\ref{disy1}, one can see that the signal can already be found directly with the basic selection criteria choice in Eq.(\ref{eq13}).

At quantum electrodynamics (QED) perturbative expansion, the main signal  $J/\psi \to 2\gamma + q \bar{q}$ 
is a next-to-next-to-leading process while the leading processes is $J/\psi \to q \bar{q}$. 
Therefore, there are double infrared divergence and 
double collinear divergence in the main signal process before adding all the virtual correction together. Although 
the two photons must be detected in the signal which can be guaranteed in the experimental measurement by $E_\gamma>0.05$. 
But the $E_\gamma$ and $cos\theta_{\gamma q, \gamma \bar{q}}$ distributions obtained in our theoretical calculation 
may far away from real one in $E_\gamma\rightarrow 0$ and $cos\theta_{\gamma q, \gamma \bar{q}}\rightarrow 1$ range, 
and the reliable results in these range should be obtained by resummation to all perturbative order. 
From the $E_\gamma$ distribution and $cos\theta_{\gamma q, \gamma \bar{q}}$ distribution in the Fig.~\ref{disy1} , 
sharp peaks are observed in these range.  Therefore, we have to put a set of cut 
as $E_\gamma>0.3$ GeV and $cos\theta_{\gamma g,\gamma q, \gamma \bar{q}}<0.85$ to avoid unreliable estimation in the following 
investigation. These cut values are just a rough consideration without detailed calculation support. 

In the following study, the basic cut values in TABLE~\ref{tab:one} are modified as:
\begin{equation}
E_{\gamma, cut}=0.30 \text{ GeV},\quad
cos\theta_{cut}=0.93,\quad
s_{cut}=
	\begin{cases}
	(2 \times 0.13 \text{ GeV})^2, & \text{for $u$ or $d$ quarks} \\ 
	(2 \times 0.49 \text{ GeV})^2, & \text{for $s$ quark} 
	\end{cases} ,\quad
cos\theta_{\gamma g, \gamma q, cut}=0.85.
\label{eq14}
\end{equation}

However, the cut values for selection criteria can be optimized in the analysis. 
We can find the best cut values to suppress the background and highlight the signal, 
and the target is to seek these cut values to maximize the ratio R. 
The best cut values can be obtained by the following strategy:
\begin{itemize}
	\item step 1: Fix all the selection criteria to the basic cut values in Eq.(\ref{eq14}).  
	\item step 2: Obtain the best cut value for a selection criterion from the cut value distribution of the ratio R 
	              by fixing other selection criteria. 
	\item step 3: Fix the best cut values from the previous criteria, and repeat step 2 until all best selection criteria are obtained.
\end{itemize}

The specific searching process is shown here:
\begin{itemize}
	\item Find the best cut value $E_{\gamma, cut}$ for criterion 1: the relation of the ratio R to $E_{\gamma, cut}$ is shown in FIG.~\ref{math1}
	with the basic cut values for criterion 2-4. The result shows that the best cut value is $E_{\gamma, cut}=0.30$ GeV.

\begin{figure}[ht]      
        \centering
        \subfloat[]{
                \includegraphics[width=8cm]{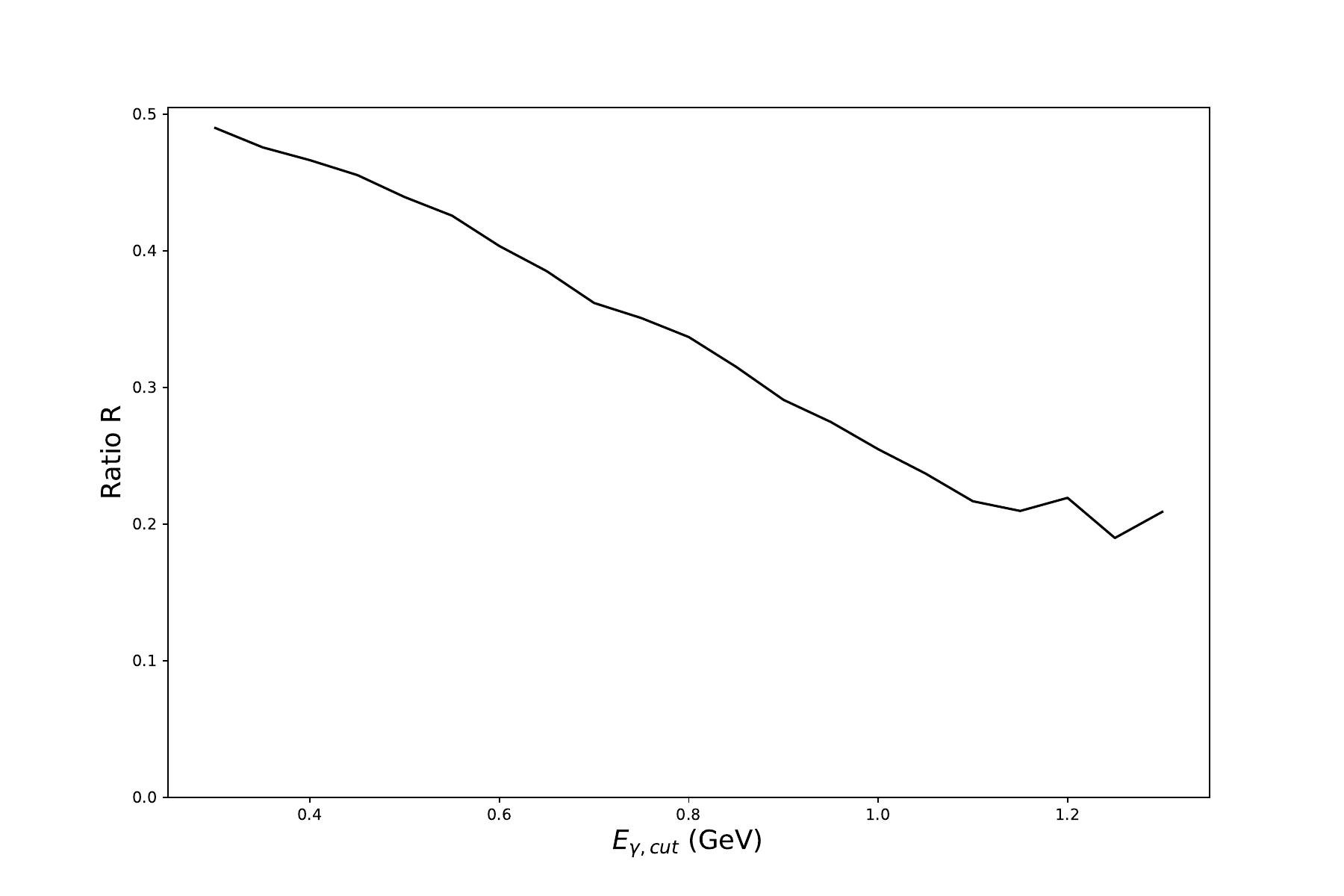}
                \label{math1}
        }
        \subfloat[]{
                \includegraphics[width=8cm]{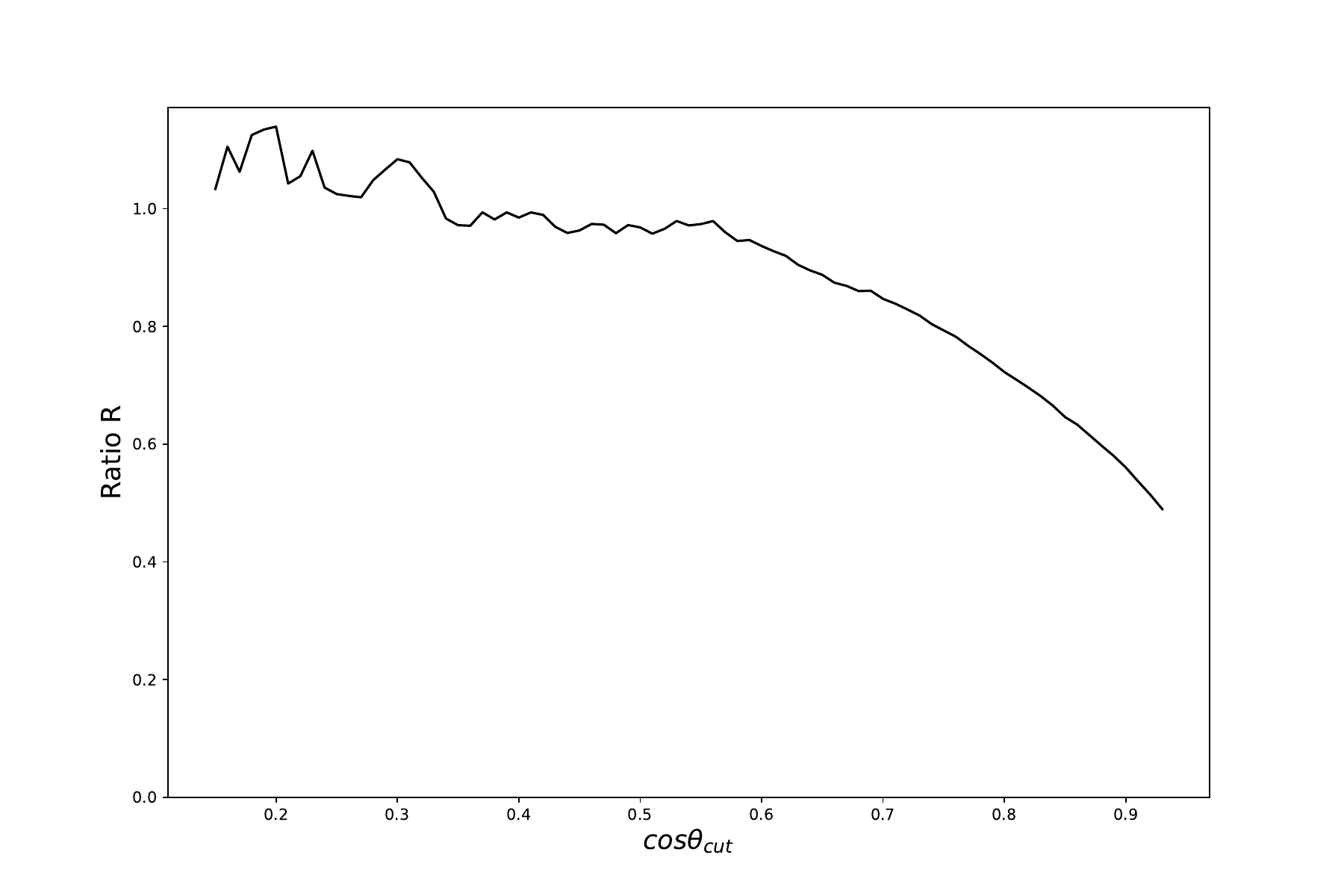}
                \label{math2}
        } \\
        \subfloat[]{
                \includegraphics[width=8cm]{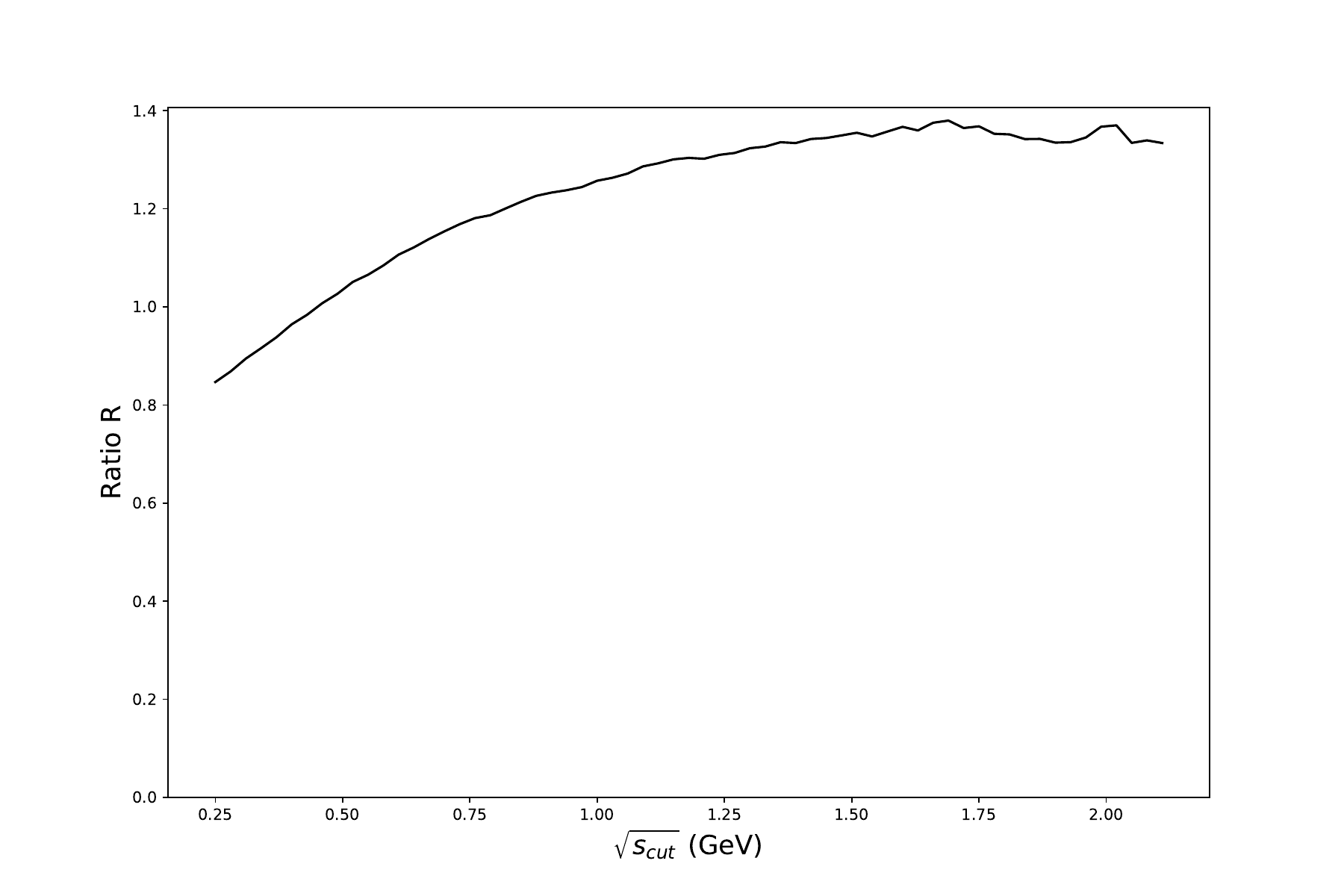}
                \label{math25}
        }
    \subfloat[]{
        \includegraphics[width=8cm]{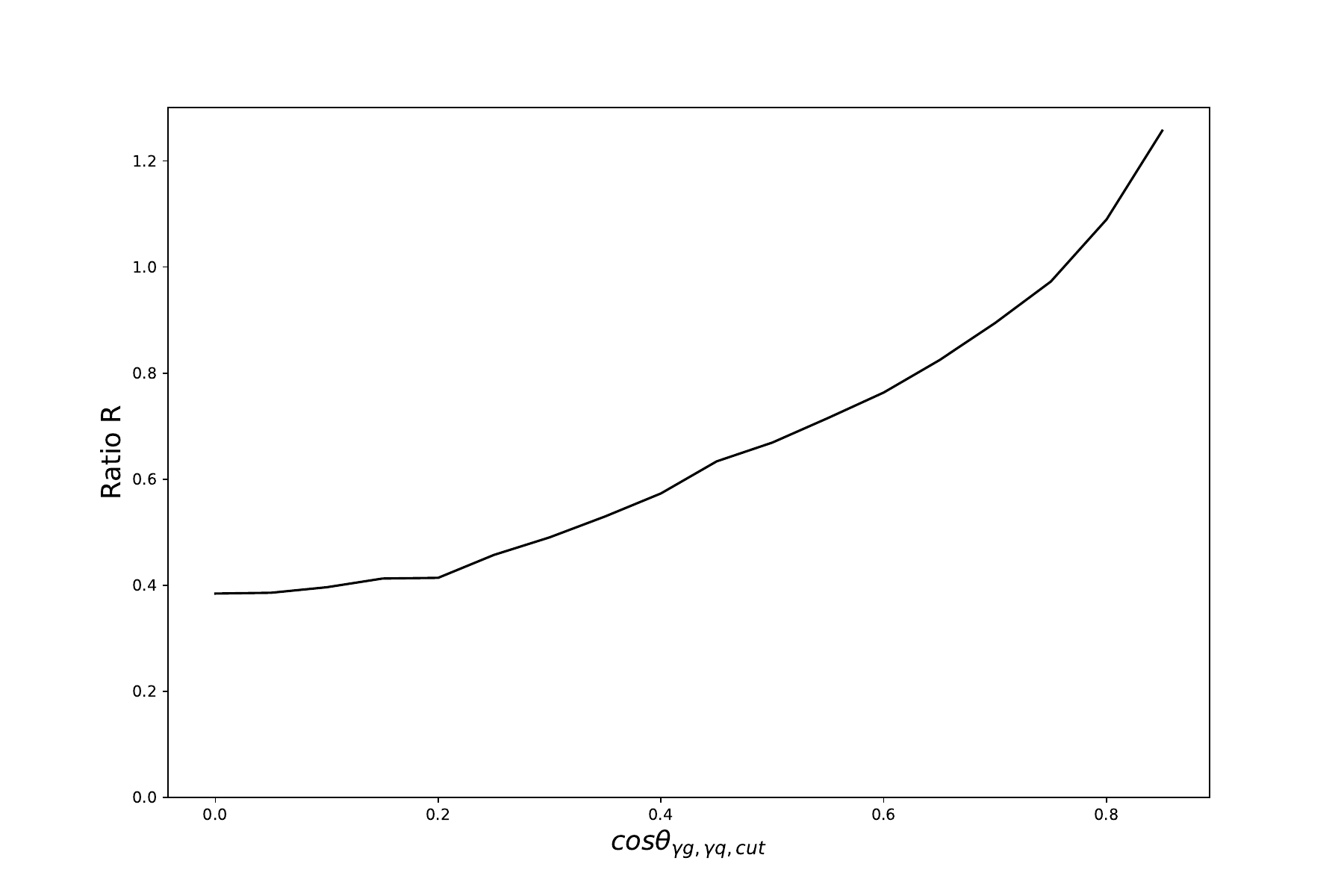}
        \label{math3}
    }
        \caption{The relation of the ratio R to
        different cut values }
        \label{rdisy1}
\end{figure}

    \item Find the best cut value $cos\theta_{cut}$ for criterion 2: the relation of the ratio R to $cos\theta_{cut}$ is shown in FIG.~\ref{math2}
    with $E_{\gamma, cut}=0.30$ GeV and the basic cut values for criterion 3-4. In this condition, the more events are cut off, the greater the ratio R. Hence we do a balance between R and the number of events, and adopt $cos\theta_{cut}=0.70$ as the best cut value.

    \item Find the best cut value $s_{cut}$ for criterion 3: the relation of the ratio R to $\sqrt{s_{cut}}$ is shown in FIG.~\ref{math25}
    with $E_{\gamma, cut}=0.30$ GeV, $cos\theta_{cut}=0.70$  and the basic cut value for criterion 4. One can see that R is only a small increase when $s_{cut}>(1.00 \text{ GeV})^2$. Hence we adopt that the best cut value is $s_{cut}=(1.00 \text{ GeV})^2$. 
	
    \item Find the best cut value for criterion 4: the relation of the ratio R to $cos\theta_{\gamma g, \gamma q, cut}$ is shown in FIG.~\ref{math3}
	with $E_{\gamma, cut}=0.30$ GeV, $cos\theta_{cut}=0.70$  and $s_{cut}=(1.00 \text{ GeV})^2$. 
	The result shows that the best cut value is $cos\theta_{\gamma g, \gamma q, cut}=0.85$.
\end{itemize}

Finally the best cut values summarized as
\begin{equation}
E_{\gamma, cut}=0.30 \text{ GeV},\quad
cos\theta_{cut}=0.70,\quad
s_{cut}=(1.00 \text{ GeV})^2 ,\quad
cos\theta_{\gamma g, \gamma q, cut}=0.85, 
\label{eq15}
\end{equation}
and the results are obtained as 
\begin{equation}
\begin{split}	
N(J/\psi \to \gamma \gamma u \bar{u})=3812,&
\quad
N(J/\psi \to \gamma \gamma d \bar{d})=71,
\quad 
N(J/\psi \to \gamma \gamma s \bar{s})=71, \\
N(J/\psi \to \gamma \gamma g g g)=928, &
\quad 
N(J/\psi \to \gamma \gamma q \bar{q})=3954, \\
N(e^{+} e^{-} \to \gamma \gamma u \bar{u})=2974,&
\quad
N(e^{+} e^{-} \to \gamma \gamma d \bar{d})=485,
\quad 
N(e^{+} e^{-} \to \gamma \gamma s \bar{s})=485, \\
N(e^{+} e^{-} \to \gamma \gamma + hadrons)=3944,& 
\quad 		
N(J/\psi \to \gamma \gamma + hadrons )=4882,
\quad
R=1.24.
\end{split}
\end{equation} 

\begin{figure}[ht]
        \centering
        \subfloat{
                \includegraphics[width=8cm]{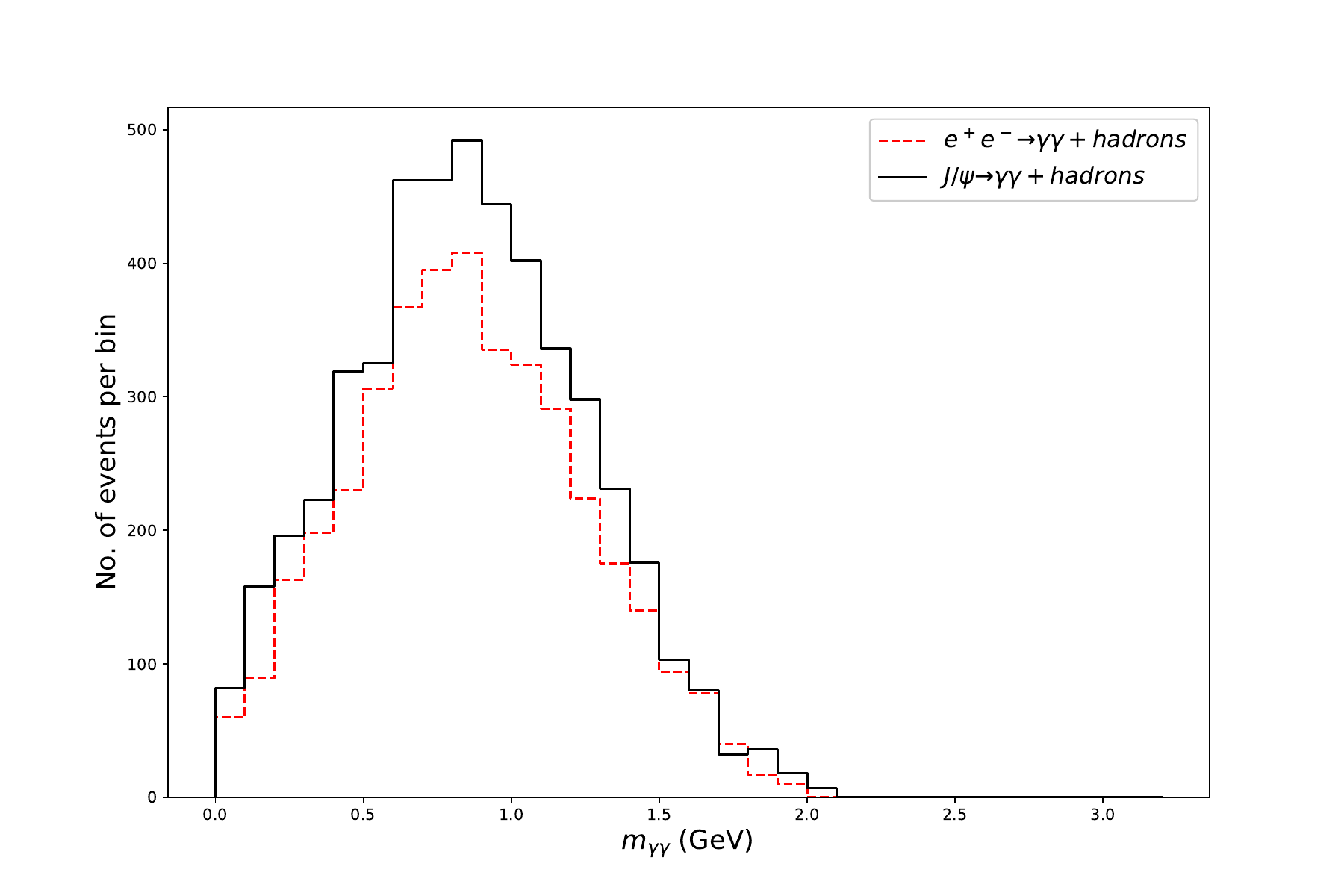}
        }
        \subfloat{
                \includegraphics[width=8cm]{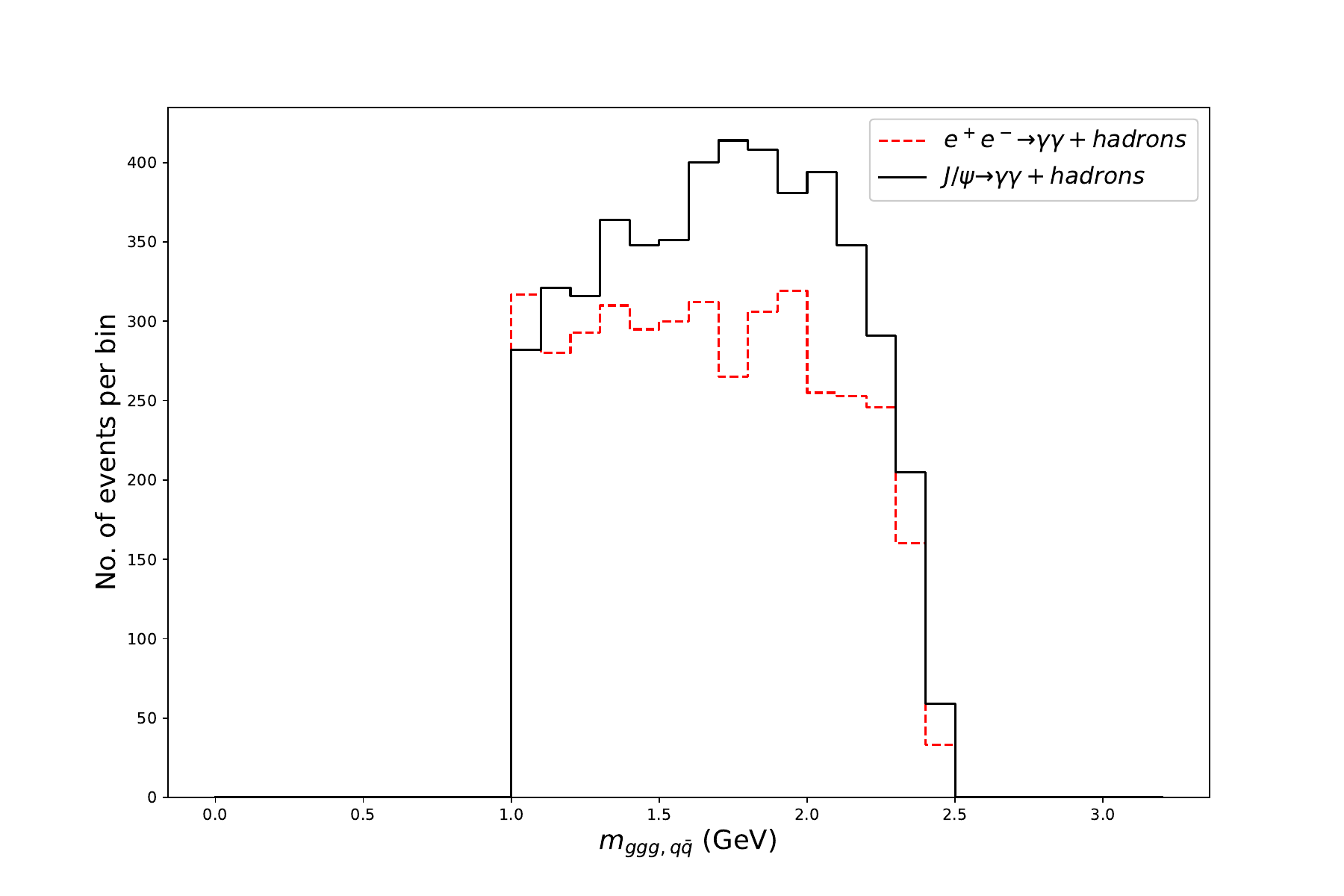}
        }\\
    \subfloat{
        \includegraphics[width=8cm]{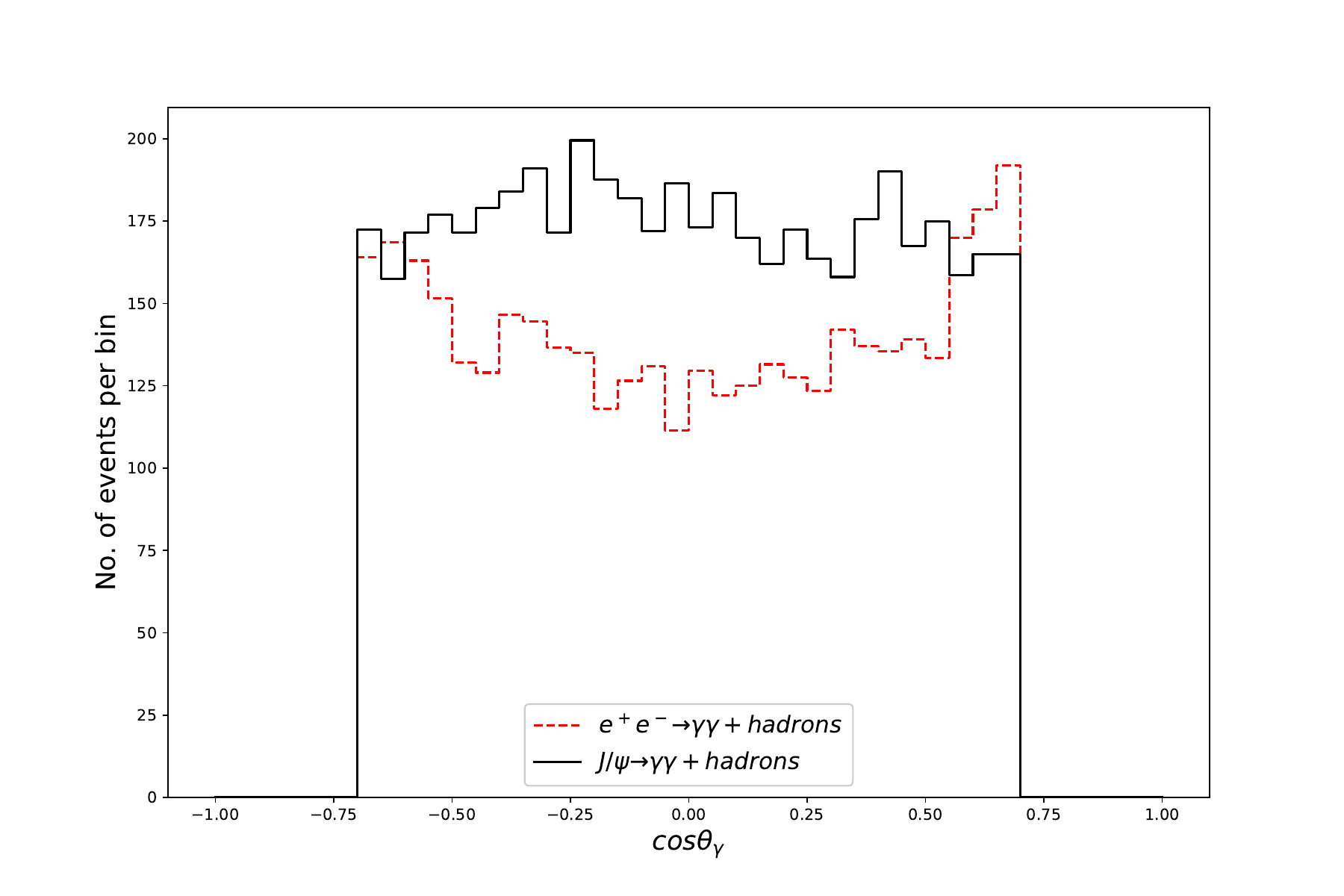}
    }
    \subfloat{
    	\includegraphics[width=8cm]{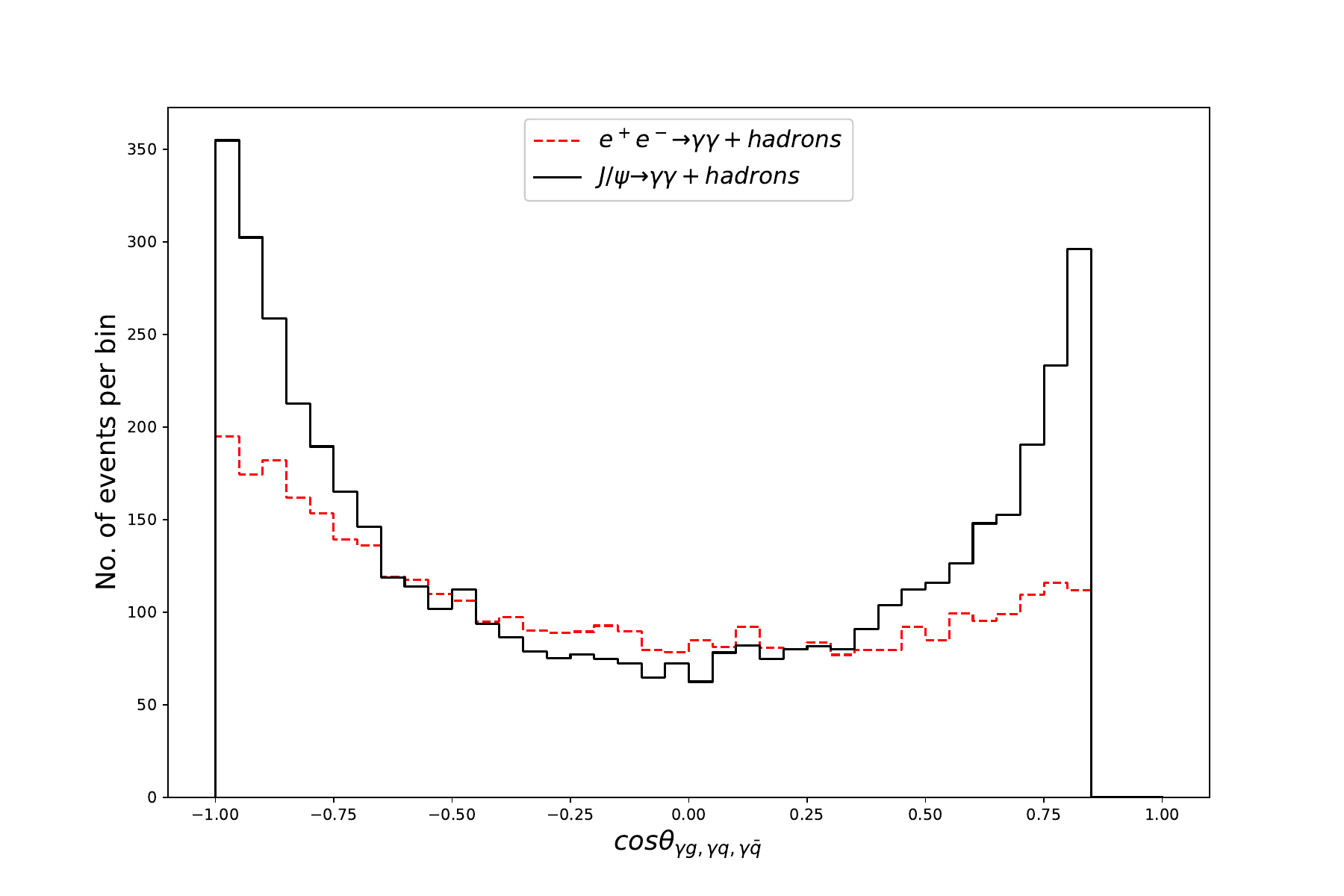}
    }\\
    \subfloat{
        \includegraphics[width=8cm]{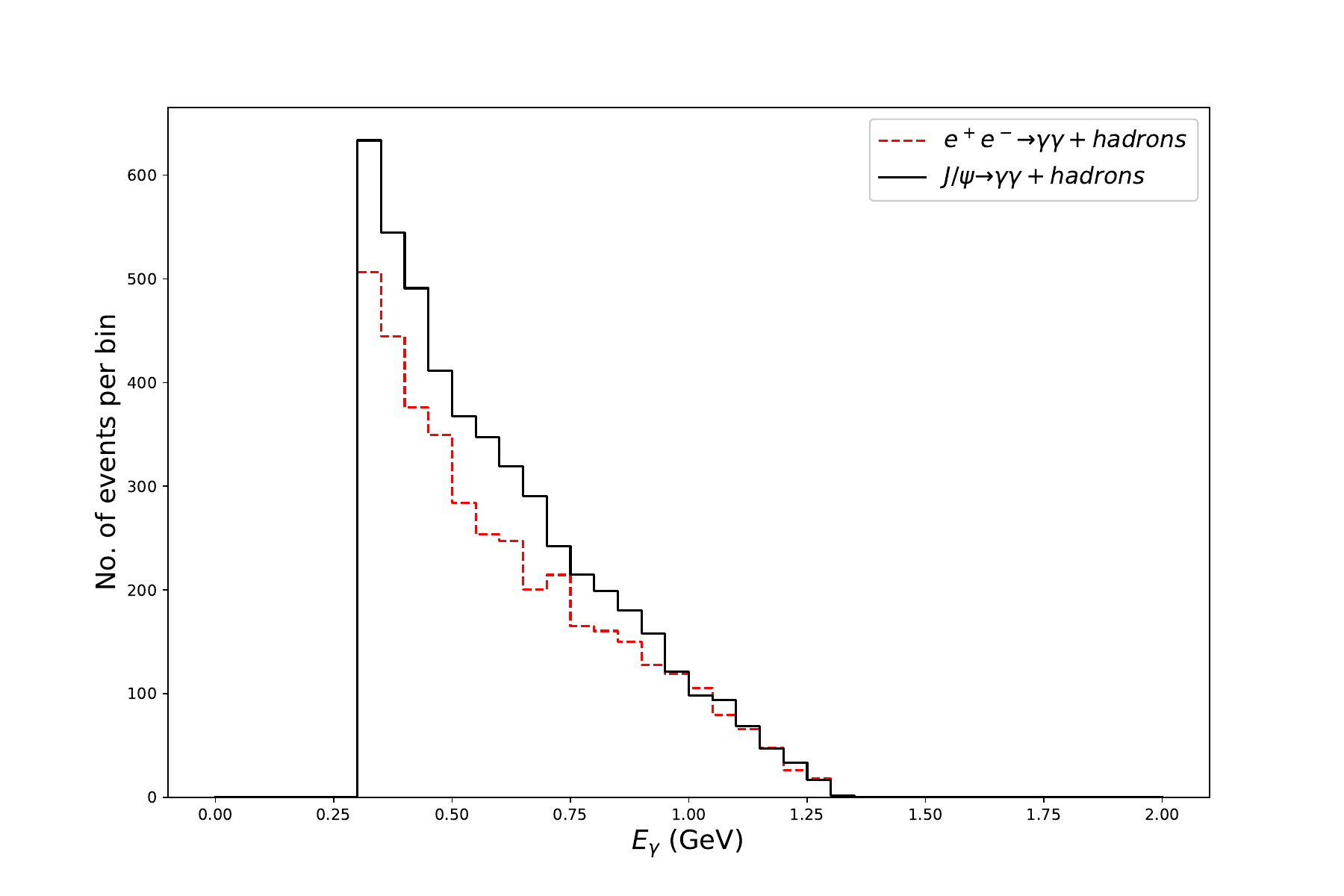}
    }
        \caption{The distributions of $J/\psi \to \gamma \gamma + hadrons$ and $e^{+} e^{-} \to \gamma \gamma + hadrons$ with the best selection criteria.
        The bin widths are 0.1 GeV for $m_{\gamma\gamma}$ and $m_{ggg, q \bar{q}}$, 0.05 GeV for $E_\gamma$, and 0.05 for $cos\theta_\gamma$ and $cos\theta_{\gamma g, \gamma q, \gamma \bar{q}}$. }
\label{dis1}
\end{figure}

With the best cut values given in Eq.(\ref{eq15}), the five distributions for the signal $J/\psi \to \gamma \gamma + hadrons$ and 
the background $e^{+} e^{-} \to \gamma \gamma + hadrons$ are shown in FIG.~\ref{dis1}. 
From these distribution, it is clear that the signal $J/\psi \to \gamma \gamma + hadrons$ is of the same order in 
magnitude as the background $e^{+} e^{-} \to \gamma \gamma + hadrons$, and 
the background may be subtracted from sideband estimation in the experiment measurement.

\section{Summary and Discussion}
In summary, two-photon radiative decay process $J/\psi \to 2\gamma+hadrons$ is studied, and the main contribution processes $J/\psi \to 2\gamma + g g g$ and  $J/\psi \to 2\gamma + q \bar{q}$ are calculated.  
For the situation of $J/\psi$ data sample at the BESIII, the signal and the main background 
$e^{+} e^{-} \to \gamma \gamma + hadrons (q \bar{q})$ are investigated. 
The best selection criteria in experiment are also investigated and the ratio of signal to background can reach 1.24
with 4882 signal events. The five distributions of the signal and background are also presented,
and the distributions show that the signal is large enough to be measured. It is expected that the measurement
on the two-photon radiative decay $J/\psi \to 2\gamma+hadrons$ could be conducted in the future. 

It should be pointed out that there are double infrared divergence and 
double collinear divergence for the main signal process $J/\psi \to 2\gamma + q \bar{q}$, 
and the $E_\gamma$ and $cos\theta_{\gamma q, \gamma \bar{q}}$ distributions obtained in our fix-order perturbative calculation 
can not be trusted in $E_\gamma\rightarrow 0$ and $cos\theta_{\gamma q, \gamma \bar{q}}\rightarrow 1$ range. 
When these distributions are obtained in future experimental measurement, further study should be performed to obtain 
the reliable results in these range by QED resummation to all perturbative order. 
However, it is out of the scope of this work. 

\begin{acknowledgments}

This work was supported by the National Natural Science Foundation of China with Grant No. 12135013, and in part by National Key Research and Development Program of China under Contract No. 2020YFA0406400.

\end{acknowledgments}

\bibliography{refs.bib}

\begin{thebibliography}{12}%
\makeatletter
\providecommand \@ifxundefined [1]{%
 \@ifx{#1\undefined}
}%
\providecommand \@ifnum [1]{%
 \ifnum #1\expandafter \@firstoftwo
 \else \expandafter \@secondoftwo
 \fi
}%
\providecommand \@ifx [1]{%
 \ifx #1\expandafter \@firstoftwo
 \else \expandafter \@secondoftwo
 \fi
}%
\providecommand \natexlab [1]{#1}%
\providecommand \enquote  [1]{``#1''}%
\providecommand \bibnamefont  [1]{#1}%
\providecommand \bibfnamefont [1]{#1}%
\providecommand \citenamefont [1]{#1}%
\providecommand \href@noop [0]{\@secondoftwo}%
\providecommand \href [0]{\begingroup \@sanitize@url \@href}%
\providecommand \@href[1]{\@@startlink{#1}\@@href}%
\providecommand \@@href[1]{\endgroup#1\@@endlink}%
\providecommand \@sanitize@url [0]{\catcode `\\12\catcode `\$12\catcode
  `\&12\catcode `\#12\catcode `\^12\catcode `\_12\catcode `\%12\relax}%
\providecommand \@@startlink[1]{}%
\providecommand \@@endlink[0]{}%
\providecommand \url  [0]{\begingroup\@sanitize@url \@url }%
\providecommand \@url [1]{\endgroup\@href {#1}{\urlprefix }}%
\providecommand \urlprefix  [0]{URL }%
\providecommand \Eprint [0]{\href }%
\providecommand \doibase [0]{https://doi.org/}%
\providecommand \selectlanguage [0]{\@gobble}%
\providecommand \bibinfo  [0]{\@secondoftwo}%
\providecommand \bibfield  [0]{\@secondoftwo}%
\providecommand \translation [1]{[#1]}%
\providecommand \BibitemOpen [0]{}%
\providecommand \bibitemStop [0]{}%
\providecommand \bibitemNoStop [0]{.\EOS\space}%
\providecommand \EOS [0]{\spacefactor3000\relax}%
\providecommand \BibitemShut  [1]{\csname bibitem#1\endcsname}%
\let\auto@bib@innerbib\@empty
\bibitem [{\citenamefont {Ablikim}\ \emph {et~al.}(2012)\citenamefont {Ablikim}
  \emph {et~al.}}]{BESIII:2012pbg}%
  \BibitemOpen
  \bibfield  {author} {\bibinfo {author} {\bibfnamefont {M.}~\bibnamefont
  {Ablikim}} \emph {et~al.} (\bibinfo {collaboration} {BESIII}),\ }\bibfield
  {title} {\bibinfo {title} {{Determination of the number of $J/\psi$ events
  with $J/\psi \rightarrow \, inclusive$ decays}},\ }\href
  {https://doi.org/10.1088/1674-1137/36/10/001} {\bibfield  {journal} {\bibinfo
   {journal} {Chin. Phys. C}\ }\textbf {\bibinfo {volume} {36}},\ \bibinfo
  {pages} {915} (\bibinfo {year} {2012})},\ \Eprint
  {https://arxiv.org/abs/1207.2865} {arXiv:1207.2865 [hep-ex]} \BibitemShut
  {NoStop}%
\bibitem [{\citenamefont {Ablikim}\ \emph {et~al.}(2017)\citenamefont {Ablikim}
  \emph {et~al.}}]{BESIII:2016kpv}%
  \BibitemOpen
  \bibfield  {author} {\bibinfo {author} {\bibfnamefont {M.}~\bibnamefont
  {Ablikim}} \emph {et~al.} (\bibinfo {collaboration} {BESIII}),\ }\bibfield
  {title} {\bibinfo {title} {{Determination of the number of $J/\psi$ events
  with inclusive $J/\psi$ decays}},\ }\href
  {https://doi.org/10.1088/1674-1137/41/1/013001} {\bibfield  {journal}
  {\bibinfo  {journal} {Chin. Phys. C}\ }\textbf {\bibinfo {volume} {41}},\
  \bibinfo {pages} {013001} (\bibinfo {year} {2017})},\ \Eprint
  {https://arxiv.org/abs/1607.00738} {arXiv:1607.00738 [hep-ex]} \BibitemShut
  {NoStop}%
\bibitem [{BES(2021)}]{BESIII:2021cxx}%
  \BibitemOpen
  \bibfield  {title} {\bibinfo {title} {{Number of $J/\psi$ events at
  BESIII}},\ }\href@noop {} {\  (\bibinfo {year} {2021})},\ \Eprint
  {https://arxiv.org/abs/2111.07571} {arXiv:2111.07571 [hep-ex]} \BibitemShut
  {NoStop}%
\bibitem [{\citenamefont {Chen}\ \emph {et~al.}(2021)\citenamefont {Chen},
  \citenamefont {Jia}, \citenamefont {Mo}, \citenamefont {Pan},\ and\
  \citenamefont {Xiong}}]{Chen:2020bju}%
  \BibitemOpen
  \bibfield  {author} {\bibinfo {author} {\bibfnamefont {W.}~\bibnamefont
  {Chen}}, \bibinfo {author} {\bibfnamefont {Y.}~\bibnamefont {Jia}}, \bibinfo
  {author} {\bibfnamefont {Z.}~\bibnamefont {Mo}}, \bibinfo {author}
  {\bibfnamefont {J.}~\bibnamefont {Pan}},\ and\ \bibinfo {author}
  {\bibfnamefont {X.}~\bibnamefont {Xiong}},\ }\bibfield  {title} {\bibinfo
  {title} {{Four-lepton decays of neutral vector mesons}},\ }\href
  {https://doi.org/10.1103/PhysRevD.104.094023} {\bibfield  {journal} {\bibinfo
   {journal} {Phys. Rev. D}\ }\textbf {\bibinfo {volume} {104}},\ \bibinfo
  {pages} {094023} (\bibinfo {year} {2021})},\ \Eprint
  {https://arxiv.org/abs/2009.12363} {arXiv:2009.12363 [hep-ph]} \BibitemShut
  {NoStop}%
\bibitem [{\citenamefont {Ablikim}\ \emph {et~al.}(2021)\citenamefont {Ablikim}
  \emph {et~al.}}]{BESIII:2021ocn}%
  \BibitemOpen
  \bibfield  {author} {\bibinfo {author} {\bibfnamefont {M.}~\bibnamefont
  {Ablikim}} \emph {et~al.} (\bibinfo {collaboration} {BESIII}),\ }\bibfield
  {title} {\bibinfo {title} {{Observation of $J/\psi$ decays to
  $e^{+}e^{-}e^{+}e^{-}$ and $e^{+}e^{-}\mu^{+}\mu^{-}$}},\ }\href@noop {} {\
  (\bibinfo {year} {2021})},\ \Eprint {https://arxiv.org/abs/2111.13881}
  {arXiv:2111.13881 [hep-ex]} \BibitemShut {NoStop}%
\bibitem [{\citenamefont {Zyla}\ \emph {et~al.}(2020)\citenamefont {Zyla} \emph
  {et~al.}}]{ParticleDataGroup:2020ssz}%
  \BibitemOpen
  \bibfield  {author} {\bibinfo {author} {\bibfnamefont {P.~A.}\ \bibnamefont
  {Zyla}} \emph {et~al.} (\bibinfo {collaboration} {Particle Data Group}),\
  }\bibfield  {title} {\bibinfo {title} {{Review of Particle Physics}},\ }\href
  {https://doi.org/10.1093/ptep/ptaa104} {\bibfield  {journal} {\bibinfo
  {journal} {PTEP}\ }\textbf {\bibinfo {volume} {2020}},\ \bibinfo {pages}
  {083C01} (\bibinfo {year} {2020})}\BibitemShut {NoStop}%
\bibitem [{\citenamefont {Coffman}\ \emph {et~al.}(1990)\citenamefont {Coffman}
  \emph {et~al.}}]{MARK-III:1989jot}%
  \BibitemOpen
  \bibfield  {author} {\bibinfo {author} {\bibfnamefont {D.}~\bibnamefont
  {Coffman}} \emph {et~al.} (\bibinfo {collaboration} {MARK-III}),\ }\bibfield
  {title} {\bibinfo {title} {{Study of the Doubly Radiative Decay $J/\psi \to
  \gamma \gamma \rho^0$}},\ }\href {https://doi.org/10.1103/PhysRevD.41.1410}
  {\bibfield  {journal} {\bibinfo  {journal} {Phys. Rev. D}\ }\textbf {\bibinfo
  {volume} {41}},\ \bibinfo {pages} {1410} (\bibinfo {year}
  {1990})}\BibitemShut {NoStop}%
\bibitem [{\citenamefont {Bai}\ \emph {et~al.}(2004)\citenamefont {Bai} \emph
  {et~al.}}]{BES:2004pec}%
  \BibitemOpen
  \bibfield  {author} {\bibinfo {author} {\bibfnamefont {J.~Z.}\ \bibnamefont
  {Bai}} \emph {et~al.} (\bibinfo {collaboration} {BES}),\ }\bibfield  {title}
  {\bibinfo {title} {{A Study of J / psi ---\ensuremath{>} gamma gamma
  V(rho,phi) decays with the BESII detector}},\ }\href
  {https://doi.org/10.1016/j.physletb.2004.04.085} {\bibfield  {journal}
  {\bibinfo  {journal} {Phys. Lett. B}\ }\textbf {\bibinfo {volume} {594}},\
  \bibinfo {pages} {47} (\bibinfo {year} {2004})},\ \Eprint
  {https://arxiv.org/abs/hep-ex/0403008} {arXiv:hep-ex/0403008} \BibitemShut
  {NoStop}%
\bibitem [{\citenamefont {Ablikim}\ \emph {et~al.}(2018)\citenamefont {Ablikim}
  \emph {et~al.}}]{BESIII:2018dim}%
  \BibitemOpen
  \bibfield  {author} {\bibinfo {author} {\bibfnamefont {M.}~\bibnamefont
  {Ablikim}} \emph {et~al.} (\bibinfo {collaboration} {BESIII}),\ }\bibfield
  {title} {\bibinfo {title} {{Study of $\eta(1475)$ and $X(1835)$ in radiative
  $J/\psi$ decays to $\gamma \phi$}},\ }\href
  {https://doi.org/10.1103/PhysRevD.97.051101} {\bibfield  {journal} {\bibinfo
  {journal} {Phys. Rev. D}\ }\textbf {\bibinfo {volume} {97}},\ \bibinfo
  {pages} {051101} (\bibinfo {year} {2018})},\ \Eprint
  {https://arxiv.org/abs/1801.02127} {arXiv:1801.02127 [hep-ex]} \BibitemShut
  {NoStop}%
\bibitem [{\citenamefont {Besson}\ \emph {et~al.}(2008)\citenamefont {Besson}
  \emph {et~al.}}]{CLEO:2008gct}%
  \BibitemOpen
  \bibfield  {author} {\bibinfo {author} {\bibfnamefont {D.}~\bibnamefont
  {Besson}} \emph {et~al.} (\bibinfo {collaboration} {CLEO}),\ }\bibfield
  {title} {\bibinfo {title} {{Inclusive Radiative J/psi Decays}},\ }\href
  {https://doi.org/10.1103/PhysRevD.78.032012} {\bibfield  {journal} {\bibinfo
  {journal} {Phys. Rev. D}\ }\textbf {\bibinfo {volume} {78}},\ \bibinfo
  {pages} {032012} (\bibinfo {year} {2008})},\ \Eprint
  {https://arxiv.org/abs/0806.0315} {arXiv:0806.0315 [hep-ex]} \BibitemShut
  {NoStop}%
\bibitem [{\citenamefont {Bodwin}\ \emph {et~al.}(1995)\citenamefont {Bodwin},
  \citenamefont {Braaten},\ and\ \citenamefont {Lepage}}]{Bodwin:1994jh}%
  \BibitemOpen
  \bibfield  {author} {\bibinfo {author} {\bibfnamefont {G.~T.}\ \bibnamefont
  {Bodwin}}, \bibinfo {author} {\bibfnamefont {E.}~\bibnamefont {Braaten}},\
  and\ \bibinfo {author} {\bibfnamefont {G.~P.}\ \bibnamefont {Lepage}},\
  }\bibfield  {title} {\bibinfo {title} {Rigorous qcd analysis of inclusive
  annihilation and production of heavy quarkonium},\ }\href@noop {} {\bibfield
  {journal} {\bibinfo  {journal} {Phys. Rev.}\ }\textbf {\bibinfo {volume}
  {D51}},\ \bibinfo {pages} {1125} (\bibinfo {year} {1995})}\BibitemShut
  {NoStop}%
\bibitem [{\citenamefont {Wang}(2004)}]{Wang:2004du}%
  \BibitemOpen
  \bibfield  {author} {\bibinfo {author} {\bibfnamefont {J.-X.}\ \bibnamefont
  {Wang}},\ }\bibfield  {title} {\bibinfo {title} {{Progress in FDC project}},\
  }\href {https://doi.org/10.1016/j.nima.2004.07.094} {\bibfield  {journal}
  {\bibinfo  {journal} {Nucl. Instrum. Meth. A}\ }\textbf {\bibinfo {volume}
  {534}},\ \bibinfo {pages} {241} (\bibinfo {year} {2004})},\ \Eprint
  {https://arxiv.org/abs/hep-ph/0407058} {arXiv:hep-ph/0407058} \BibitemShut
  {NoStop}%
\end{thebibliography}%
\end{document}